\documentclass{svproc}
\usepackage{graphicx}
\usepackage{cite}
\usepackage{url}

\begin{document}
\mainmatter              
\title{Influencing the Influencers: Evaluating Person-to-Person Influence on Social Networks Using Granger Causality \thanks{The research for this paper was supported in part by the Knight Foundation and the Office of Naval Research Grant (N000141812106) and an Omar N. Bradley Fellowship, and by the center for Informed Democracy and Social-cybersecurity  (IDeaS) and the center for Computational Analysis of Social and Organizational Systems (CASOS) at Carnegie Mellon University. The views and conclusions  are those of the authors and should not be interpreted as representing the official  policies, either expressed or implied, of the Knight Foundation, Office of Naval Research or the US Government.}}
\titlerunning{Richard Kuzma et al.}  
%
\author{Richard Kuzma, Iain J. Cruickshank, \and Kathleen M. Carley}
\authorrunning{Richard Kuzma et al.} 
\institute{CASOS Institute, Carnegie Mellon University,\\
    \{rkuzma2, icruicks\}@andrew.cmu.edu,  \\ 
    kathleen.carley@cs.cmu.edu \\
    http://www.casos.cs.cmu.edu/
        }

\maketitle              

\begin{abstract}
We introduce a novel method for analyzing person-to-person content influence on Twitter. Using an Ego-Alter framework and Granger Causality, we examine President Donald Trump (the Ego) and the people he retweets (Alters) as a case study. We find that each Alter has a different scope of influence across multiple topics, different magnitude of influence on a given topic, and the magnitude of a single Alter’s influence can vary across topics. This work is novel in its focus on person-to-person influence and content-based influence. Its impact is two-fold: (1) identifying “canaries in the coal mine” who could be observed by misinformation researchers or platforms to identify misinformation narratives before super-influencers spread them to large audiences, and (2) enabling digital marketing targeted toward upstream Alters of super-influencers.

\end{abstract}

%
\section{Introduction}
Influence in social networks remains one of the mainstays of social network analysis. Recently, with the advent of online social networks and the influence that online misinformation and disinformation can have, social influence has taken on added importance and dimensions. In this work, we build upon the foundations of social influence by examining social influence at the Ego network level. Using an Ego-Alter framework we identify an Ego (Donald Trump), examine which Twitter accounts he retweets most to identify key Alters (e.g. Ivanka Trump, Linsdey Graham), identify tweet topics from the Ego using unsupervised machine learning, build a supervised machine learning classifier to classify tweets of Alters into those topics \cite{arnaboldi2012analysis, arnaboldi2013ego}. Then for a given topic, we measured the influence of each Alter on the Ego by measuring the Granger causality on the Alter’s tweet time series on the Ego’s tweet time series.

We devised this method in order to answer the question, “who influences the influencers?”  Given the role that social media super-influencers may play in the spread of (mis)information online, identifying who influences them, what topics they are influenced on, and how much they are influenced could be key to future misinformation studies.

\begin{figure}[h]
    \centering
    \includegraphics[width=1\textwidth]{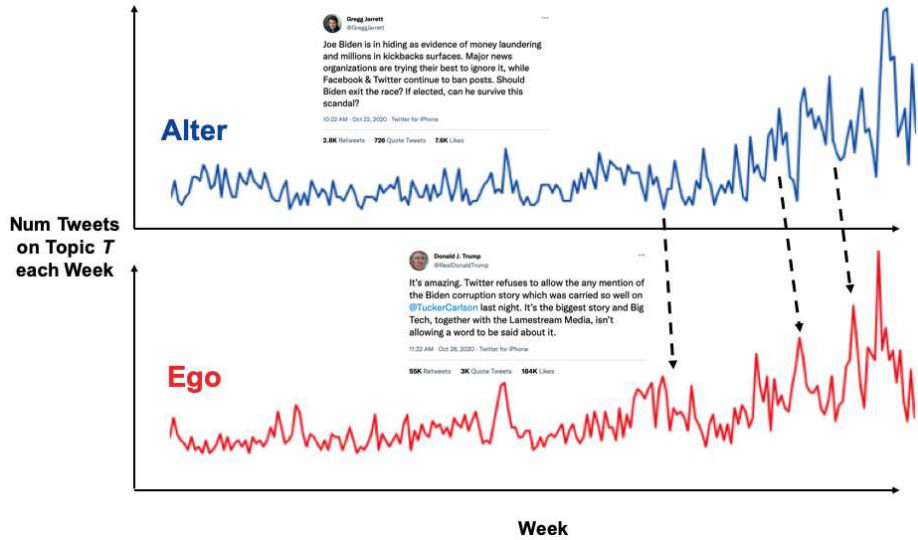}
    \caption{Example of an Alter's Tweets affecting the Ego's Tweets for a given topic. Certain topics in the Alter's tweets are then Tweeted, with a time-delay by the Ego.}
    \label{fig:example_granger}
\end{figure}

Our work is novel from other applications of Granger causality for a few reasons. Most Granger causality research on social media focuses on (1) population-level influence, (ours focuses on person-to-person influence); (2) sentiment analysis, (ours focuses on topics analysis); and (3) impact of social media behavior on real world events such as financial markets or protests (our work focuses on influence within online behavior on Twitter) \cite{bolllen2010twitter, mittal2011stock, bastos2015}. Other work using Granger causality to examine person-to-person influence on social media examined sharing of explicit URLs between Twitter accounts, but not the content or topic of tweets as we do \cite{steeg2012information}. 

%
\section{Background}
The last few years have seen incredible advances in, and need for, social cybersecurity. Social Cybersecurity is an emerging field that looks at the intersection of human behavior in a cyber mediated information environment \cite{carley2020social}. In particular, recent events like the COVID-19 pandemic and use of online disinformation in various national elections have highlighted the need to better understand the ways in which mis- and disinformation are spread online \cite{nelson2020danger, brangham2021tech}. Previous works have generally focused on the role of bots and trolls in their roles in spreading mis- and disinformation \cite{beskow2020finding, bessi2016social, Hui2019, ha2021bots, carley2020social, uyheng2021characterizing}. More recently, coordination among cyborg or fully human accounts has been studied as means by which malicious actors can propagate mis- and disinformation \cite{giglietto2020it, pacheco2020unveiling, pacheco2020uncovering}. The insight of coordinated spread by social media accounts is even being used to investigate cross-platform mis- and disinformation spread \cite{starbird2020crossplatform}. To date, much of the research in online mis- and disinformation campaigns has focused on particular types of accounts and coordination among accounts.

One area of the spread of online mis- and disinformation that is less well-studied is the role of online social media influencers. Social media influencers, or those online accounts with the ability to reach hundreds of thousands, if not millions, of other social media users with their content represent an important component of the social media landscape, by virtue of their reach. Recent work has shown that social media influencers can often have an outsized impact on the information quality of any given online discussion. For example, one study showed that the former U.S. President Donald Trump was responsible for over 30\% of the COVID-19 misinformation on Twitter \cite{evanega2020et}. As another example, the ‘dirty dozen’, a collection of only twelve accounts, are responsible for the vast majority of anti-vaccination content online \cite{dozencenter}. Additionally, recent disinformation campaigns, like Secondary Infektion have deliberately sought to target social media influencers to get them to spread disinformation \cite{2021secondary, foundation2020disinformation}. Thus it is clear that social media influencers can play a tremendous role in the spread of mis- and disinformation.

It is also important to note that the study of influence in social networks has a long history with much research into the topic. The study of influence in social networks pre-dates online social networks and has been used to understand everything from individuals’ behaviors to collective action of groups \cite{coussikorbel1995ontherelation, Friedkin321, friedkin2011social}. Many of the models designed for understanding social influence work at the adjacency matrix-level of the social networks, and are fundamentally matrix based models \cite{olshevsky2019graphtheoretic, flache2017models}. As such, these models tend to look more at population levels of influence. Less studied within the social network analysis community is the role and exercise of social influence of Ego networks.

An important consideration when studying influence is the role of causality. Granger causality is a statistical method that examines whether \(X\) has predictive power in describing \(Y\) \cite{granger1969investigating}. \(X_t\) is said to Granger-cause \(Y_t\) if \(Y_{t+1}\) can be better explained by past values of \(Y_t\) and past values of \(X_t\) than by the presence of past values of \(Y_t\) alone. 

%
\subsection{Data and Definitions}
We retrieved tweets from Donald Trump from the Trump Twitter Archive \cite{brendan2021trump}. This repository contains 56,571 tweets and retweets from Donald Trump’s twitter account. Reports indicate that not all tweets from the @RealDonaldTrump twitter account come directly from Donald Trump. Prior to the presidency, it was believed that tweets from an Android device were from Donald Trump while other platforms (iPhone, web) were not. Once Donald Trump became president, he switched to a specially-issued iPhone. At least one staffer was trusted with @RealDonaldTrump twitter privileges. Tweets came from Donald Trump, direct dictations to staffers, and presumably from the close, trusted staffers themselves attempting to channel Donald Trump’s head space \cite{chute2019he}. Given the importance of the Twitter audience to Donald Trump, we believe that all tweets from the account, even those from staffers instead of Trump himself, accurately reflect Donald Trump’s wishes and are worth studying as if they came from Trump himself.

%
\section{Methods}
\begin{figure}[h]
    \centering
    \includegraphics[width=1\textwidth]{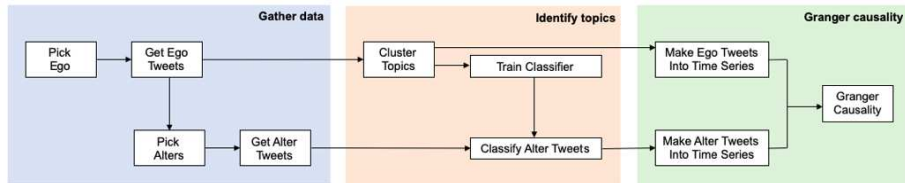}
    \caption{Overview of Method. Our method consists of three, main steps: gathering the data and constructing the ego-later network, finding topics in the textual data and building a classifier of those topics, and then testing for causality in the ego-alter network on those topics.}
    \label{fig:method_overview}
\end{figure}

Our method has three stages: gathering data, identifying topics of conversation, and determining granger causality. 

\subsection{Gathering Data}
We gathered tweets from the Trump Twitter Archive, which retained copies of all of Trump’s tweets and retweets from before Trump’s Twitter suspension. CSV and JSON files with all captured tweets are available at the Trump Twitter Archive. We limited tweets to the time that Donald Trump was President of the United States and active on Twitter, January 20th, 2017 to January 8th, 2020. We hypothesized that the information Trump ingested affected the information he output. In other words, what he retweeted affected what he tweeted. We found 12  accounts from individuals Trump retweeted the most (the Alters). We excluded Trump’s retweets of @WhiteHouse and of himself (@RealDonaldTrump). These Alters included family members, conservative media pundits, and members of Congress. 
We used SNScrape, an open-source tool, to get the tweet IDs of these 12 Alter accounts \cite{snscrape}. The original tweets of these accounts were available, but data was not available on who these Alters retweeted. We used these tweet IDs to get the tweet content and metadata through a process known as hydration, using Hydrator, another open-source tool \cite{Hydrator:2020}.

\subsection{Identifying Topics}
We cleaned Trump’s tweets by extracting URLs, mentions, and “RT” (used for manual retweets when Trump quoted another twitter user) from tweets. We also used the gensim Python library to preprocess tweets by removing punctuation and making all words lowercase. We then tokenized the tweets and found common bigrams and trigrams using the gensim library. Bigrams are two-word sequences like “crooked hillary” or “sleepy joe” and trigrams are three-word sequences like “make america great.” Following tokenization, we embedded the tweets using the Universal Sentence Encoder (USE) from Google Research which was available at the TensorFlow Hub \cite{ceruniversal, USE}. We also featurized all tweets using the term frequency-inverse document frequency (TF-IDF) from the scikit learn library \cite{scikit-learn}.

We chose the USE because it was built for short sequences ranging from a sentence to a short paragraph in length, which is similar to the length of a tweet, and performs well in semantic similarity tasks, determining if one sequence is similar to another. We performed k-medoids clustering \cite{velmurugan:2020} on the USE embeddings to identify topic clusters and within those topic groups examined the most important words by averaging the TF-IDF scores of all documents within a given USE cluster. 

We chose to cluster only on the Ego's tweets because including Alter tweets in the topic classifier would dilute the Ego’s tweet content with Alter content when forming clusters. With 12 Alters in our case study, it is possible topics would drift toward Alter content rather than Ego content, making a classifier predict whether a given Alter’s tweet was like other Alter tweets rather than similar to the Ego’s tweet. We settled on eight topics for Trump’s tweets through a combination of silhouette score \cite{Shahapure:2020} from the k-medoid clustering of USE embeddings and interpretation of the TF-IDF top words for each cluster. 

\begin{figure}[h]
    \centering
    \includegraphics[width=1\textwidth]{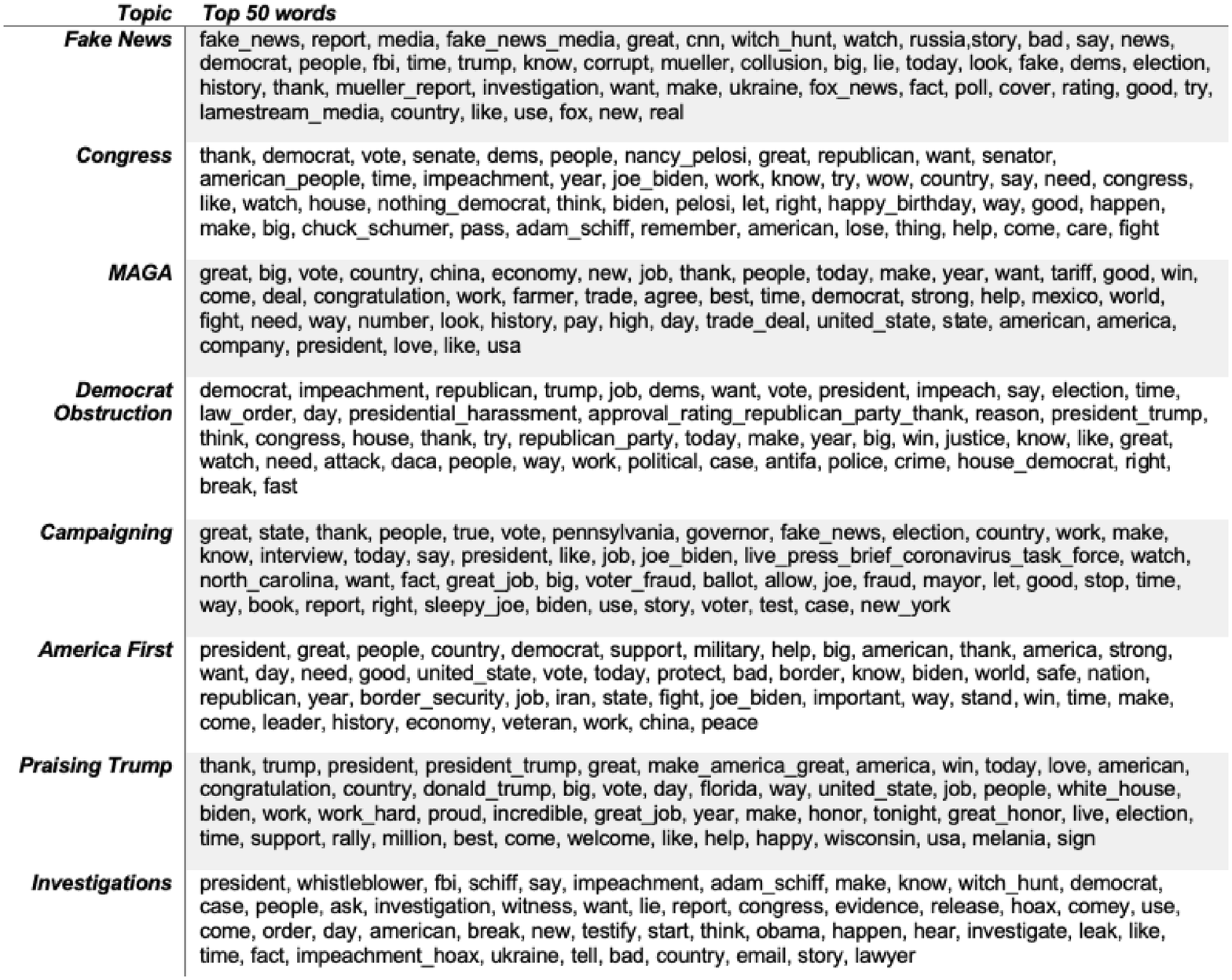}
    \caption{Top 50 words and phrases used by Donald Trump in each topic (ranked by TF-IDF score). Clusters determined by USE embeddings.}
    \label{fig:top_words}
\end{figure}

We categorized each Alter’s tweets into one of the categories of Trump’s tweets. To do this we train a Support Vector Classifier on Trump’s tweets, using the USE embeddings as features and the unsupervised topic clusters as the label. Our final model had a weighted F-1 score of 0.91. 

For each Alter, we clean the tweets using the same cleaning pipeline as Trump’s tweets then embed the tweets using USE embeddings. We then use the model trained to predict Trump’s tweet categories to predict the Alter’s tweet categories. Now that the Alter’s tweets are categorized, we turn them into eight time series, one for each of the topic clusters. We use weeks as time-steps because we found days contained too few tweets to make an effective time series for each topic.

\subsection{Granger Causality}
We create a time-series with units of weeks on the x-axis and number of user’s tweets on a given topic on the y-axis. The Ego, \(E\), and  Alters, \(A_{1}\) to \(A_{n}\), each each have \(m\) time series, where \(m\) is the number of topics. Then, for a given Alter (e.g. Ivanka Trump) and a given topic (e.g. Campaigns) we want to find if the Alter’s tweets granger-cause (have predictive power to describe) the the tweets of the Ego (e.g. Donald Trump).

We first considered granger causality at the level of days, but found that most users did not have enough data when viewing topics at the day level. We think weeks are a viable time scale because intuition suggests that an Alter may tweet about a topic of importance to them repeatedly. In future work, it would be worth considering a 2-3 day rolling window

Testing for granger causality requires that both time series are stationary, that is, the mean, variance, and autocorrelation do not change with time. Stationary was checked with the Augmented Dickey Fuller test \cite{greene:1997}. If a time series is not stationary, it may be made stationary by differencing the time series. This requires subtracting the number of tweets at time $t$ from the number of tweets at time $t-1$. If the time series of the Alter’s or Trump’s tweets for a given topic are not stationary, both time series were differenced and stationarity re-examined. This way both time series always had the same amount of differencing.

If the time series could be made stationary, a granger causality test can then  be done. Again, this test determines if \(Y_t\) (Trump’s tweets on a topic) is better explained by \(Y_{t-n}\) (Trump’s previous tweets on that topic) alone or \(Y_{t-n}\) and \(X_{t-n}\) (Alter’s previous tweets on that same topic). 

%
\section{Results}
Different Alters  influence the Ego on different topics. Even among Alters whose tweets granger-cause Trump’s tweets, some Alters have a more immediate effect (1 week) while others affect Trump’s tweets with six to eight weeks of lag.

Members of Congress (Mark Meadows, Jim Jordan, GOP Chairwoman) had influence on the “Congress” topic with one week lag between their tweets and Trump’s. Jim Jordan and GOP Chairwoman also influenced the “Investigations” topic with a one week lag. 

Conservative pundits (Mark Levin, Gregg Jarrett, and Tom Fitton) influenced the most topics (6-7), though not all with immediate effect. Their tweets on “Fake News,” “Campaigning,”, and “Investigations” (along with pundit Charlie Kirk, though less so with Tom Fitton) had 1-3 week lags between their tweets and Trumps. 

Donald Trump Jr had influence on “Fake News,” “Congress,” “Campaigning,” and “America First” topics, all with 1-3 week lags. Ivanka Trump only influenced “America First” and “Praising Trump” but with 5-6 week lags. Eric Trump had influence on more topics, but a mix of immediate lag (1-3 week) and delayed lag (6-7 week) influence.

Among Alters that influenced five or more topics (Eric Trump, Mark LEvin, Gregg Jarrett, and Tom Fitton), the immediacy of influence was not consistent. Often the Alter had a few topics with immediate lag influence (1-2 week) and a few with delayed lag influence (7-8 week).

We chose to do individual hypothesis testing and felt the low p-values would not substantively change with multiple hypothesis testing, but future work may choose to multiple hypothesis testing for added rigor.

\begin{figure}[h]
    \centering
    \includegraphics[width=1\textwidth]{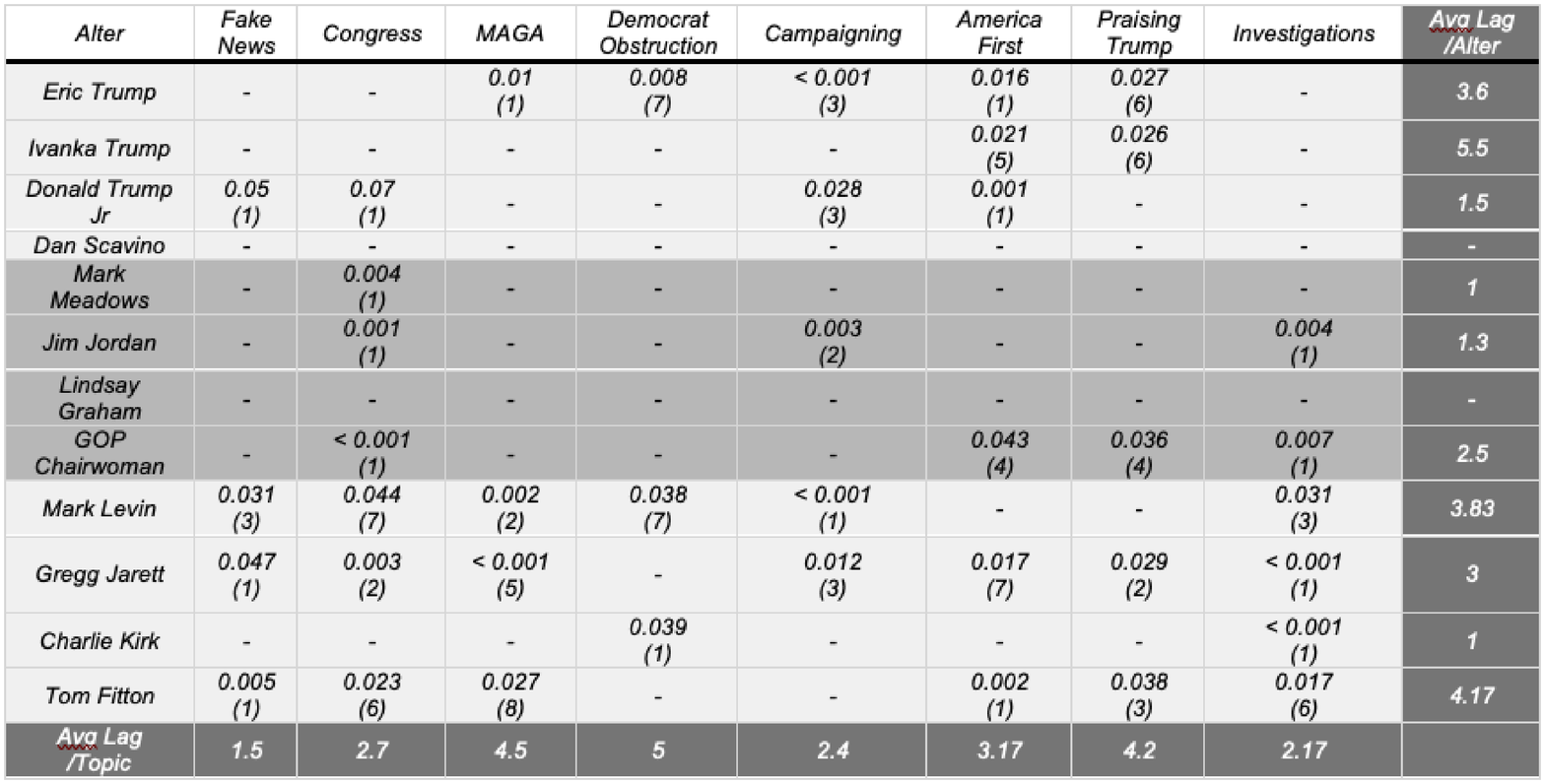}
    \caption{P-Values for granger causality test. Num lags in parenthesis (max 8 weeks). Alters shown in three groups: (1) family/close advisors, (2) members of Congress, and (3) political commentators.}
    \label{fig:granger_results}
\end{figure}

%
%
\section{Discussion}
In this paper we ask “who influences the influencers?” and introduce a novel method for examining person-to-person influence on Twitter using an Ego-Alter framework for finding which Alters influence the Ego’s posts on social media, on which topics these Alters influence the Ego, and the time delay between Alter posts and Ego posts. We show this specifically with tweets from Donald Trump and the first-order Alters he interacts with on Twitter, but believe this work could be extended to examine other influencers on Twitter and their networks, to include second and third-order Alters farther away from the Ego’s immediate network, and attempted on other social media platforms (Parler, Facebook).

Our results show substantial variation in the influence of Alters on an Ego. While our work confirms the general theories of social influence in that social influence affects an individual's behavior, we also note what influences an individual and when they influence that individual are heterogeneous entities. Some Alters have a greater scope of influence. Of Donald Trump’s eight tweet topics, some Alters influence only one or two topics while others influence seven of the eight topics. Some Alters have narrow, but powerful, influence. Charlie Kirk, a conservative talk show host, influences Trump on only two topics, “investigations” and “democrat obstruction.” Trump’s tweets lag Kirk’s by only one week, showing that Kirk has high influence on these topics. Five other Alters influence the “investigations” topic and three of them also have a one-week lag on their influence, so Kirk is not special in this category. On “democrat obstruction,” however, only two other Alters with influence. Kirk has a one-week lag between his “democrat obstruction” tweets and Trumps, while the other Alters have a seven-week lag, indicating less influence on this topic.

Different Alters influence different topics and should not be viewed as substitutable. Tom Fitton, a conservative activist and President of Judicial Watch, is very influential in the “fake news,” “america first,” and “praising trump” categories whereas Jim Jordan, a Congressman from Ohio, is highly influential on topics of “congress,” “campaigning,” and “investigations.”

Some Alters influence different topics differently. This suggests that some Alters are more influential on certain topics than others. For instance, Gregg Jarrett, a conservative American television commentator, influences Trump’s tweets on seven of eight topics. On “fake news” and “investigations” his tweets influence Trump’s after a one-week lag. On “MAGA” and “America First” his tweets influence Trump’s after a five-week and seven-week lag, respectively. Jarett would be a good influence vector for “fake news” and “investigations” but not as good for “MAGA” and “America First”

\subsection{Future Work}
Despite the useful results and demonstration of using granger-causality in understanding peer-to-peer influence on social media, there are limitations to this work. We examine a specific case of influence surrounding Donald Trump, without any comparison to how these Alters affected other Twitter users or how other Egos are influenced by their respective Alters . We focused on this base case because the Trump Twitter Archive permitted us to study the retweets of the Ego to identify the most-retweeted Alters. Generally, retweets of specific Twitter accounts are not available through the Twitter API nor are they available through Python libraries like SNScrape, making data collection difficult. If retweets of Alters were available, research could be done on whether second-order Alters (e.g. someone who Ivanka Trump retweets a lot) has influence on the Ego (Donald Trump) through the first-order Alter (Ivanka Trump).

We see two uses of this influence work. First, in the disinformation space, using Alters as “canaries in the coal mine” who could be observed by misinformation researchers or platforms to notice new misinformation narratives before super-influencers catch hold of these narratives and spread them to millions of people. Second, in the marketing space, advertisers may not have the budget to pay a super-influencer to market their product, but by paying the Alters of the Ego, the advertiser may be able to influence the super-influencer to support the product being marketed.


%
%
\bibliographystyle{splncs03}
\bibliography{refs}

\end{document}